\documentclass[12pt,preprint]{aastex}

\begin{document}

\title{Spectroscopy from Photometry Using Sparsity.\\
The SDSS Case Study}
\author{A. Asensio Ramos \and C. Allende Prieto}

\affil{Instituto de Astrof\'{\i}sica de Canarias, 38205, La Laguna, Tenerife, Spain}
\email{aasensio@iac.es}

\begin{abstract}
We explore whether medium-resolution stellar spectra can be reconstructed
from photometric observations, taking advantage of the highly compressible nature of the
spectra. We formulate the spectral reconstruction as
a least-squares problem with a sparsity constraint. In our test case using data from the
Sloan Digital Sky Survey,  only three broad-band filters are used as input.
We demonstrate that reconstruction using  three
principal components is feasible with these filters, leading to differences  with respect
to the original spectrum smaller than 5\%. We analyze the effect of uncertainties in the observed magnitudes and find
that the available high photometric precision induces very small errors in the reconstruction.
This process may facilitate the extraction of purely spectroscopic
quantities, such as the overall metallicity, for hundreds of millions of stars for which
only photometric information is available, using standard techniques applied
to the reconstructed spectra.
\end{abstract}




\section{Introduction}
There are more than 200 photometric systems that have been used in astronomy
\citep[][and references therein]{bessell05}.  The amount of
information about a star that can be extracted from photometry
is highly dependent on the choice of photometric filters.
Early on, Str\"omgren introduced the $uvby$ system with the aim of characterizing the main stellar
atmospheric parameters and reddening \citep{stromgren51,stromgren56}. Str\"omgren's
filters have widths of the order of 200 \AA, and are thus considered
intermediate-band. Other systems with similar and narrower passbands
have been introduced since, but most
photometric systems use filters significantly broader than Stromgren's,
and therefore tend to provide lower sensitivity to the atmospheric parameters.

Until recently, the most widely-used photometric system was the broad-band
Johnson-Cousins $UBVRI$, but with the advent of the Sloan Digital
Sky Survey (SDSS), which includes CCD-based photometry for 357
million unique optical sources over more than 11,000 square degrees
\citep{abazajian09}, and 2MASS \citep{skrutskie06}, which includes
nearly half a million near-IR sources over the entire sky, these new
systems have taken over. The hegemony of the SDSS system in
the optical is illustrated by the fact that new and future instruments,
such as the Large Synoptic Survey Telescope (LSST)\footnote{See http://www.lsst.org},
the Dark Energy Survey (DES) camera\footnote{See http://www.darkenergysurvey.org}, 
or OSIRIS \citep{cepa00} on the Gran Telescopio Canarias (GTC)
are adopting the same system.

Despite the widths of the SDSS filters are 3--6 times larger than the
Str\"omgren passbands, \cite{ivezic08} have shown that, if the 
reddening is known with sufficient accuracy, it is possible to estimate stellar effective
temperatures for late-type stars with a typical precision of $\sim 100$ K,
and metallicities with a precision of $\sim 0.2$ dex from SDSS
photometry alone for stars of moderate metallicities.
The sensitivity of SDSS photometry to surface
gravity is much weaker, and disentangling this parameter
from the other two, even in the absence of reddening, may not be
possible, but in any case we are not aware of any sucessful calibration.

Because of the vast number of sources with available SDSS photometry, it is
desirable to ensure that we are extracting all possible information
captured by this system. The most straightforward techniques for mapping
photometric indices into the quantities of interest, such as stellar
atmospheric parameters, have provided limited success, and the time
is ripe to explore new avenues. This work examines the possibility of using a simple technique
inspired on the concept of compressed sensing \citep[CS;][]{candes06,donoho06} to reconstruct
spectroscopic data of stellar objects from SDSS photometry. If there
is an accessible mapping between photometry and
intermediate-dispersion spectra, the available tools for
spectroscopic analysis could then be applied to the reconstructed
spectra in order to recover the parameters of interest.

The recent work on SDSS spectra by \cite{mcgurk10}
indicates that for stars in narrow color bands
(about 0.02 mag wide in $g-r$), principal component analysis (PCA)
can be applied, and most of the variance  in the spectra is
recovered with just 4 principal components. This result strongly
suggests that the SDSS intermediate-resolution spectra are highly
sparse, and the most relevant information in the data can be
compressed into just four numbers.
This statement should be accompanied by a warning: SDSS spectra typically
have a modest signal-to-noise ratio (most SDSS
spectra are for stars in the range $16<g<18.6$ and have typical
signal-to-noise ratios at 500 nm between 65 and 8).
We also note that \cite{mcgurk10} used the median difference between
the original and reconstructed spectra to quantify the level of
agreement. Hence, significantly larger differences between the 
original and the spectra recovered
with four components are expected for a small fraction of their sample, as
we show in this paper.

If the spectra are indeed sparse, there is a good chance that
photometry alone can be used to reconstruct them to some precision 
\citep[see][for similar applications]{asensio_lopez_cs10,asensio_an10}.
In this paper, we explore this possibility in detail, in
particular for the case when SDSS stellar spectra can be represented
as a linear combination of a small number of vectors, as concluded 
by \cite{mcgurk10}. Section 2 describes the basic concept and
develops the mathematical method. Section 3 applies the method
to SDSS spectra and photometry. Section 4 discusses the error propagation from the photometry to the
reconstructed spectra, and Section 5 summarizes our findings.

\section{Sparsity and reconstruction}
During the last few years, the emerging theory of compressed sensing 
is showing that the Nyquist-Shannon sampling theorem
is too restrictive in case some details of the signal structure are known in advance. 
The interesting point of the new CS paradigm is that, in many instances, natural signals have a structure
that is known in advance. The key point is that, typically, only few elements of the
basis set in which we develop the signal are necessary for an accurate description
of the important physical information. Instead of measuring the full signal (wavelength variation of the stellar
spectrum in our case), under the CS framework one measures a few linear projections of the signal
along some vectors known in advance and reconstructs the signal solving a non-linear problem.

Explicity, let $\mathbf{f}$ be a vector of length $M$ that represents the
sampled wavelength variation of the stellar spectrum. A standard spectrograph measures
the spectrum by accumulating photons in wavelength bins determined by the spectral
resolution. Instead, we propose to measure scalar products of the signal with
carefully selected vectors (multiplex measurements), so that:
\begin{equation}
\mathbf{y} = \mathbf{\Phi} \mathbf{f} + \mathbf{e},
\label{eq:sensing}
\end{equation}
where $\mathbf{y}$ is the vector of measurements of dimension $N$, $\mathbf{\Phi}$ is 
an $N\times M$ sensing matrix that we particularize below and
$\mathbf{e}$ is a vector of dimension $N$ that characterizes the noise on the measurement process.
Note that the previous equation describes the most general linear multiplexing scheme in which the number of
measurements $N$ and the length of the signal $M$ may differ. In the most standard
multiplexing situation, the number of scalar products measured equals the dimension
of the signal ($N=M$) and it is possible to recover the vector $\mathbf{f}$ provided
that $\mathrm{rank}(\mathbf{\Phi})=N$ (in other words, that every row of the $\mathbf{\Phi}$ matrix 
is orthogonal with respect to every other row).

Our aim, though, is to solve the previous linear system (i.e., obtain the spectrum $\mathbf{f}$) from
the smallest possible number of measurements $\mathbf{y}$. In general, this can be accomplished
by solving the linear system using the singular value decomposition (SVD) of the $\mathbf{\Phi}$
matrix \citep[see, e.g.,][]{numerical_recipes86}. The solution through the SVD fulfills that it is the one producing the smallest
$\ell_2$-norm\footnote{The $\ell_n$-norm of a vector is
given by $\parallel \mathbf{x} \parallel_n = (\sum_i |x_i|^n)^{1/n}$ if $n > 0$. The $\ell_0$ pseudo-norm 
is given by the number of non-zero elements of $\mathbf{x}$.} of the residuals, or equivalently, the 
least-squares solution. When $N \ll M$,
this solution is strongly affected by noise and is practically useless in general.

However, this problem can be overcome if the ingredient of sparsity is invoked. The success
of CS techniques is fundamentally based on the idea that, if the signal of interest is 
sparse in a certain basis set (or can be efficiently compressed in this basis set), the
reconstruction is made possible. Any compressible signal\footnote{A signal is said to be compressible (or quasi-sparse) if 
it is possible to find a basis for which the projection coefficients along the vectors
of the basis reordered in
decreasing magnitude decay in absolute value like a power-law.} can be written, in general, in the following way:
\begin{equation}
\mathbf{f} = \mathbf{W}^\dag \mathbf{a},
\label{eq:basis_decomposition}
\end{equation}
where now $\mathbf{a}$ is a $K$-sparse\footnote{A vector is $K$-sparse if only $K$ elements of
the vector are different from zero.} vector of size $M$ and $\mathbf{W}^\dag$ is the
transpose of an $M \times M$ transformation matrix associated with the basis set in which 
the signal is sparse. For instance, $\mathbf{W}$ can be the Fourier matrix if the signal
$\mathbf{f}$ is the combination of a few sinusoidal components. In our case, we will use
the transformation matrix associated with the principal components.

The combination of the previous two ingredients leads to the following multiplexing scheme:
\begin{equation}
\mathbf{y} = \mathbf{\Phi} \mathbf{W}^\dag \mathbf{a} + \mathbf{e},
\label{eq:sensing2}
\end{equation}
with the hypothesis that $\mathbf{a}$ is sparse, i.e., that the $\ell_0$-norm
of $\mathbf{a}$ is as small as possible.

\subsection{Sparsity}
Principal component analysis \citep[PCA;][]{pearson01,karhunen47,loeve55} has been 
applied to SDSS data with the
aim of classification, noise-reduction and compression \citep[e.g.,][]{connolly95,yip04}
for different types of objects. The principal components (eigenvectors
of the correlation/covariance matrix of the database) represent
a complete basis set for a given database of spectra.


One of the advantages of the PCA decomposition is that the 
importance of an eigenvector (measured as the associated absolute value of the eigenvalue)
decays typically like a power-law. Therefore, if one considers that only $M$ principal components contribute 
significantly to the
reconstruction of a spectrum, it can be seen that the vector $\mathbf{a}$ in 
Eq. (\ref{eq:basis_decomposition}) is non-zero only in the first $M$ elements, and approximately zero
in the rest. Additionally, the matrix $\mathbf{W}^\dag$ is built from the principal components
ordered from the absolute value of their associated eigenvalues as columns.

Recently, \cite{mcgurk10} have applied PCA to SDSS stellar
spectra. They have analyzed a subset of the full spectral database of SDSS and calculated the 
principal components separately for stars in intervals of 0.02 mag in the $g-r$ color. The range
of colors considered spans $-0.2 < g-r < 0.9$, corresponding to MK spectral types A3 to K3. According
to \cite{ivezic05}, this segregation in $g-r$ color is roughly equivalent to a segregation in effective temperature
due to the large correlation between this parameter and the $g-r$ color. It is also of interest to point
out that the effect of reddening is limited by selecting stars with an estimated extinction below 0.3 mag in the $r$ band.
Thanks to the binning, the number of principal components needed in each interval to reach 
noise level is highly reduced. They demonstrate that the mean spectrum plus three principal 
components (hereafter referred to as the first four principal components) are more than enough 
to statistically reconstruct the stellar spectra at the noise level. 

As a caveat, note that the quality of the principal component decomposition of \cite{mcgurk10} is only
measured through the median difference. It is then expected that $\sim$50\% of
the stars in each bin have a decomposition that reproduce the spectra with a difference
larger than the noise level (see \S\ref{sec:reconstruction}). If a different binning is proposed in the future leading to
new (hopefully improved) principal components, our reconstruction scheme remains unchanged and 
can be computed using exactly the same observations. Thankfully, the segregation of \cite{ivezic05}
is done using an observed quantity and the bin can be known just using photometric data. We take this 
highly efficient PCA decomposition for reconstructing SDSS stellar spectra
from photometric measurements.

\subsection{Sensing matrix}
\begin{figure}[!t]
\centering
\includegraphics[width=\columnwidth]{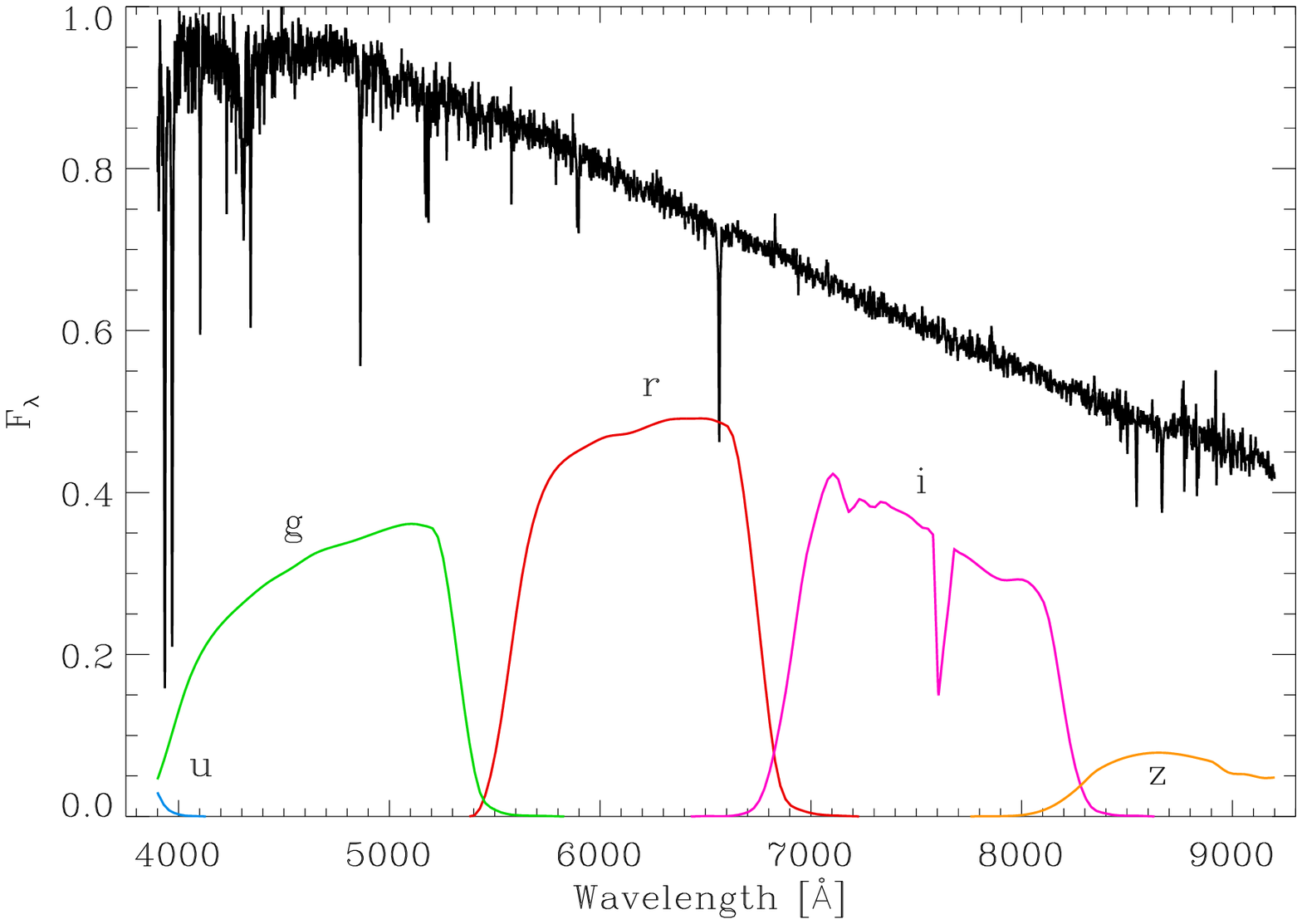}
\caption{Sample observed spectrum of an F-type star with $[\mathrm{Fe}/\mathrm{H}] \simeq -1.6$, $T_\mathrm{eff} \simeq 5886$ K
and $\log g \simeq 4.61$. The vertical axis is in flux units normalized to the maximum in the observed
spectral region. We also show the total efficiency including atmospheric transmission
for the SDSS filter set at 1.3 air masses for point-like objects. Note that filters
$u$ and $z$ have contributions outside from the observed spectral region, so that
they cannot be used for reconstruction.}
\label{fig:sensing_matrix}
\end{figure}

In addition to the sparsity condition, the other important ingredient of our technique relies on
the election of the sensing matrix. This matrix is the one that relates the underlying spectrum
with the measurements we use in the reconstruction. Our aim is to test whether photometric data
can be used to reconstruct spectra, so that the sensing matrix is not a choice, but
given by the weighting
functions of the SDSS filter set. Figure \ref{fig:sensing_matrix} shows the $ugriz$ filter
set and an example of an observed spectrum. Note that filters $u$ and $z$ have important contributions
outside the observed spectral range. Consequently, the information they contain 
cannot be easily utilized under the scheme presented in this paper and we carry out the reconstructions using 
only filters $g$, $r$ and $i$.

Obtaining the magnitude in the filter $k$ of the SDSS system from spectroscopic data 
reduces to the calculation of the
following quantity \citep{fukugita96}:
\begin{equation}
m_k = -2.5 \log_{10} \frac{\int f(\nu) S_k(\nu) \mathrm{d}\ln \nu}{\int S_k(\nu) \mathrm{d}\ln \nu}-48.6,
\label{eq:magnitudes}
\end{equation}
where $f(\nu)$ is the flux distribution of the star.
Therefore, $S(\nu)$ is the effective
transmissivity of the filter, including the filter response, the CCD quantum
efficiency and the typical transmission of the sky for a point source (in
our cases adopted for an airmass of 1.3\footnote{http://www.sdss.org/dr7/instruments/imager/filters/index.html}).
The flux measured for each filter can be estimated using the Riemann integral
as (note that more precise quadrature rules can be used without a significant change
in the following discussion):
\begin{equation}
\int f(\nu) S_k(\nu) \mathrm{d}\ln \nu \approx \sum_{i=1}^N S_k(\nu_i) f(\lambda_i) \frac{\lambda_i}{c} \Delta \lambda_i,
\end{equation}
where we have made the integration in the wavelength axis and used $f(\lambda)$ instead of
$f(\nu)$ since the principal components of \cite{mcgurk10} are given in terms of $f(\lambda)$. The normalization constant for 
each filter is obtained likewise:
\begin{equation}
C_k = \int S_k(\nu) \mathrm{d}\ln \nu \approx \sum_{i=1}^N S_k(\nu_i) \frac{\Delta \lambda_i}{\lambda_i}.
\end{equation}

Note that one is able to isolate the flux from the observed magnitudes as:
\begin{equation}
10^{-0.4(m_k+48.6)} \approx \frac{1}{C_k} \sum_{i=1}^N f(\lambda_i) 
S_k(\nu_i) \frac{\lambda_i}{c} \Delta \lambda_i.
\label{eq:measured_flux}
\end{equation}
Consequently, the flux associated to the measured photometric quantity can be written as the
dot product of the original flux distribution $f(\lambda_i)$ and a weighting function, so that each
column of the sensing matrix $\mathbf{\Phi}$ in Eq. (\ref{eq:sensing}) is given by:
\begin{equation}
\Phi_{ik} = \frac{S_k(\nu_i)}{C_k} \frac{\lambda_i}{c} \Delta \lambda_i.
\label{eq:sensing_matrix_final}
\end{equation}

\subsection{Reconstruction}
\label{sec:reconstruction}
The sparsity constraint of the spectrum is fulfilled automatically when using a
principal component decomposition, with the additional advantage of knowing exactly which coefficients of the
sparse vector $\mathbf{a}$ are non-zero. Therefore,
the solution of the problem given by Eq. (\ref{eq:sensing2}) is simpler
than the full CS problem in which the non-zero elements of $\mathbf{a}$
have to be identified. Consequently, the solution to Eq. (\ref{eq:sensing2}) is given by 
the sparse vector that minimizes the following $\ell_2$-norm:
\begin{equation}
\parallel \mathbf{y} - \mathbf{\Phi} \mathbf{W}^\dag \mathbf{a} \parallel_2.
\label{eq:l2norm}
\end{equation}
In other words, we look for the sparse vector $\mathbf{a}$ with the first $K$ elements different from
zero and the rest set to zero that minimizes the square difference between the 
photometric flux on the $g$, $r$ and $i$ filters and the ones reconstructed
using the previous formalism, where the flux is obtained as a linear combination of $K$
principal components. We now develop in more detail the steps to be followed.

Assume that the signal of interest $F(\lambda)$ can be written 
as a linear combination of $K$ (sparsity) PCA basis functions $B_i(\lambda)$, so that:
\begin{equation}
F(\lambda_i) = \sum_{k=1}^K a_k B_k(\lambda_i) \qquad \forall i=1,\ldots,M.
\label{eq:develop}
\end{equation}
The measurement process produces the following linear combinations:
\begin{equation}
y_j = \sum_{i=1}^{M} \Phi_{ij} F(\lambda_i) \qquad \forall j=1,\ldots,N,
\label{eq:sensing_linear_combination}
\end{equation}
where $M$ is the number of wavelength points, $N$ is the number of measurements and the 
$\Phi_{ij}$ are the matrix elements
of the sensing matrix $\mathbf{\Phi}$, given in Eq. (\ref{eq:sensing_matrix_final}). 
Plugging Eq. (\ref{eq:develop}) into Eq. (\ref{eq:sensing_linear_combination}), we end up with:
\begin{equation}
y_j = \sum_{i=1}^{M} \Phi_{ij} \sum_{k=1}^K a_k B_k(\lambda_i) = 
\sum_{k=1}^K a_k \sum_{i=1}^{M} \Phi_{ij} B_k(\lambda_i),
\end{equation}
where, making the substitution $t_k^j = \sum_{i=1}^{M} \Phi_{ij} B_k(\lambda_i)$, can
be written as:
\begin{equation}
y_j = \sum_{k=1}^K a_k t_k^j.
\end{equation}
The minimization of Eq. (\ref{eq:l2norm}) can be easily
done calculating the derivatives with respect to each $a_k$ and equating them to zero. In other words, given
the vector $\mathbf{o}$ of length $N$ with the observations (photometry), 
we define the metric function:
\begin{equation}
\chi^2 = \sum_{j=1}^{N} \frac{1}{\sigma^2_{o_j}} \left[ o_j - \sum_{k=1}^K a_k t_k^j \right]^2,
\end{equation}
where $\sigma^2_{o_j}$ is the variance associated to $o_j$ and discussed
in \S\ref{sec:influence_errors}. Then, we end up with the following set of linear equations for $a_k$:
\begin{equation}
\sum_{k=1}^K a_k \sum_{j=1}^{N} \frac{t_k^j t_l^j}{\sigma^2_{o_j}} = \sum_{j=1}^{N} \frac{o_j}{\sigma^2_{o_j}} t_l^j \qquad \forall l=1,\ldots,K
\label{eq:linear_system}
\end{equation}
In matrix form, we have:
\begin{equation}
\mathbf{G} \mathbf{a} = \mathbf{b},
\label{eq:linear_system_reconstruction}
\end{equation}
where
\begin{eqnarray}
\label{eq:G_b}
G_{kl} &=& \sum_{j=1}^{N} \frac{t_k^j t_l^j}{\sigma^2_{o_j}} \nonumber \\
b_l &=& \sum_{j=1}^{N} \frac{o_j}{\sigma^2_{o_j}} t_l^j.
\end{eqnarray}
The $t$'s are defined from the principal
components of the spectra and the photometric system --they
are common for all objects. The only additional information
needed for each object is the photometry and its
expected uncertainties. Each spectrum is reconstructed by 
computing the $\mathbf{b}$ vector, calculating
$\mathbf{G}^{-1} \mathbf{b}$, and using these numbers in the linear 
combination of Eq. (\ref{eq:develop}).

It is also of interest to note that the solution to Eq. (\ref{eq:l2norm}) can alternatively be easily carried out using
the SVD of the matrix $\mathbf{\Phi} \mathbf{W}^\dag$. We have verified empirically that this matrix of size 
$N \times M$ fulfills that $\mathrm{rank}(\mathbf{\Phi} \mathbf{W}^\dag) = N$. Thus, in order to reconstruct the
spectrum using $N$ principal components, we need to have, at least, $N$ measurements. Since we have only available the 
$g$, $r$ and $i$ magnitudes, we cannot expect to reconstruct the spectrum using the four principal components 
tabulated by \cite{mcgurk10} from only 3 measurements. As a consequence, we have to limit the reconstruction
to only the average spectrum plus two principal components, leading to slightly larger 
errors.

Summarizing, from the knowledge of the system response at each wavelength, the principal
components associated to the $g-r$ bin of the star and the $g$, $r$, and $i$ magnitudes, one is
able to reconstruct the stellar spectrum by solving the linear system of Eq. (\ref{eq:linear_system_reconstruction})
and using the coefficients in the linear expansion given by Eq. (\ref{eq:develop}).


\section{Demonstration of the technique}
\label{sec:demonstration}
\begin{figure*}[!t]
\centering
\includegraphics[width=0.49\columnwidth]{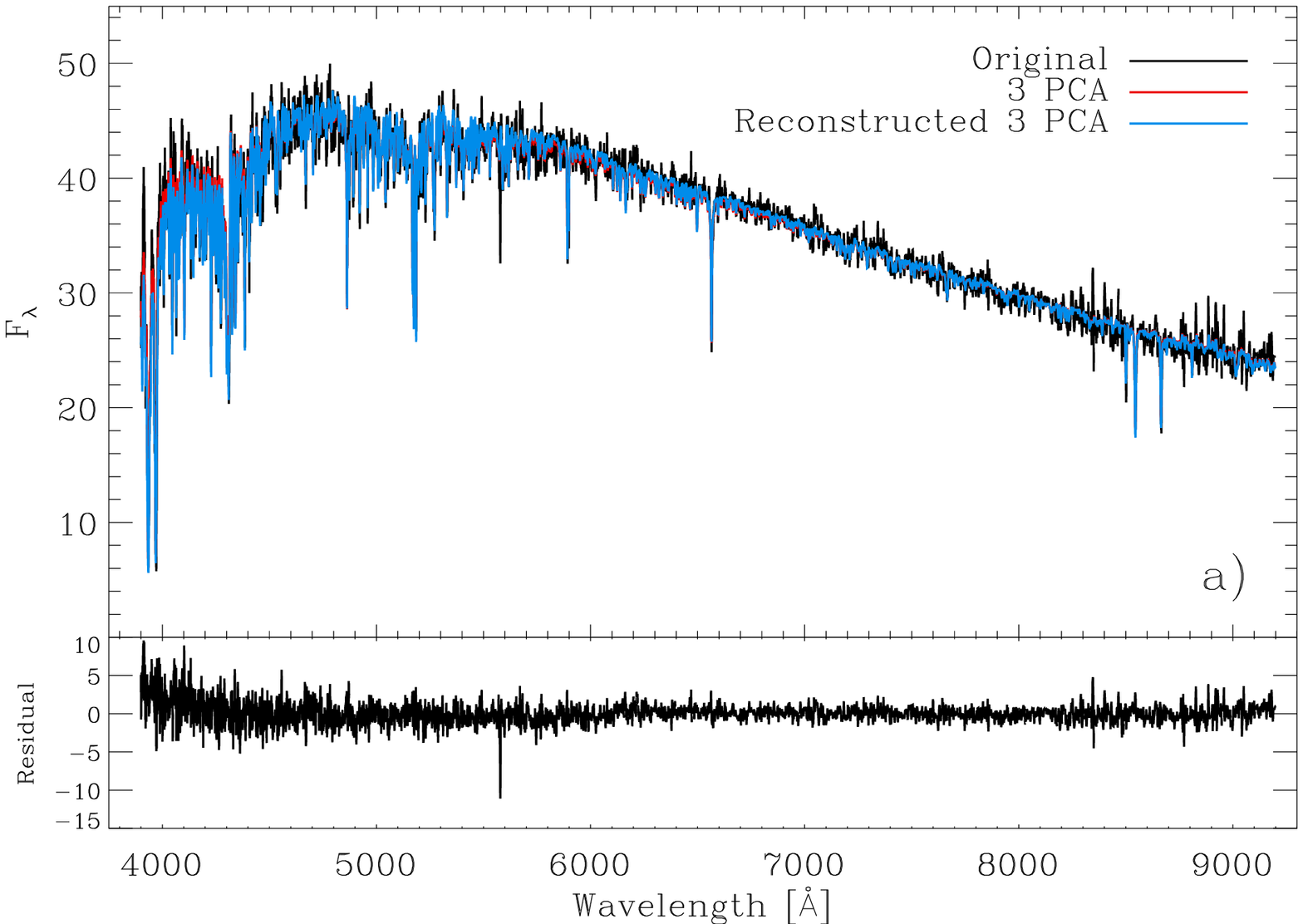}
\includegraphics[width=0.49\columnwidth]{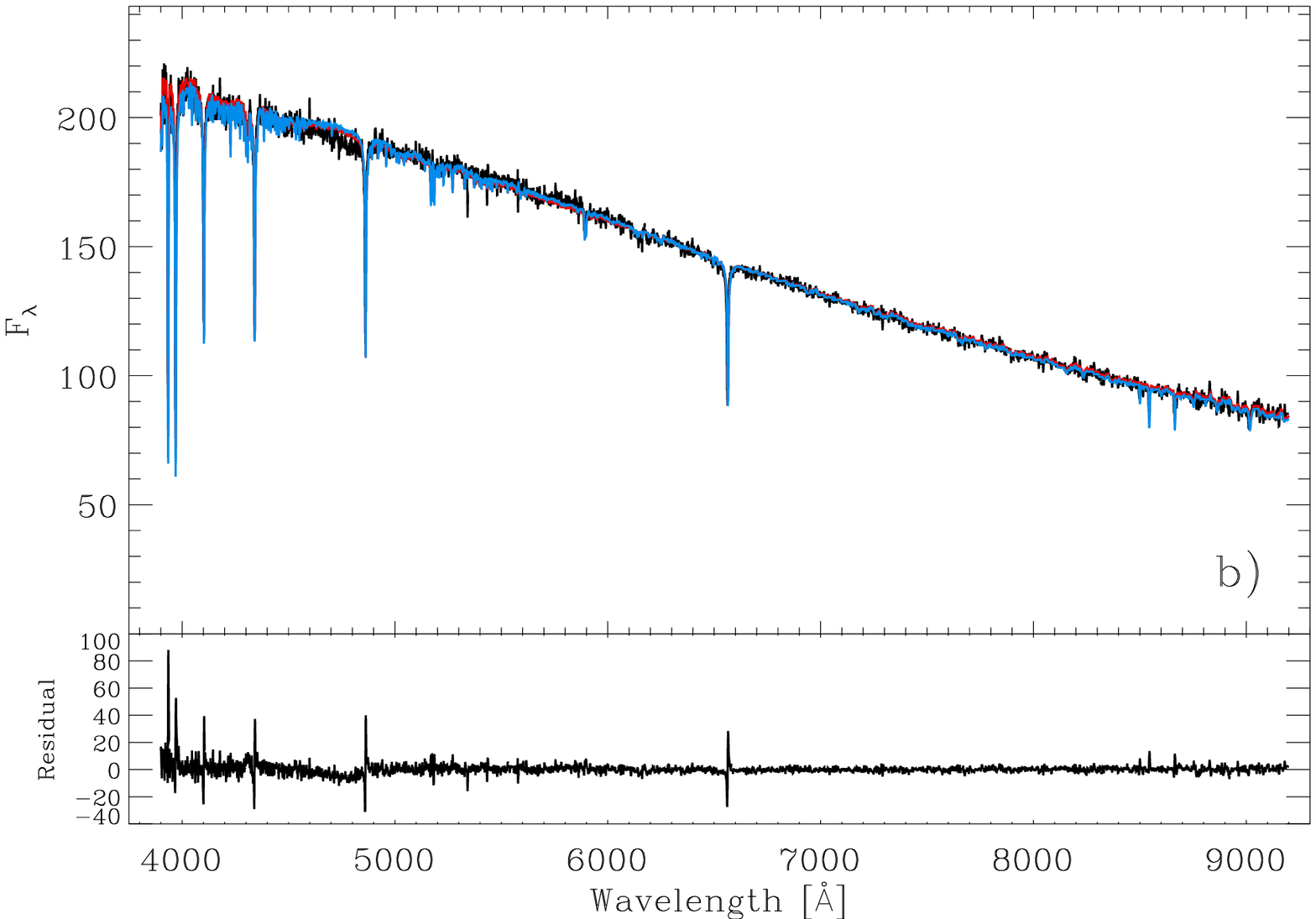}
\includegraphics[width=0.49\columnwidth]{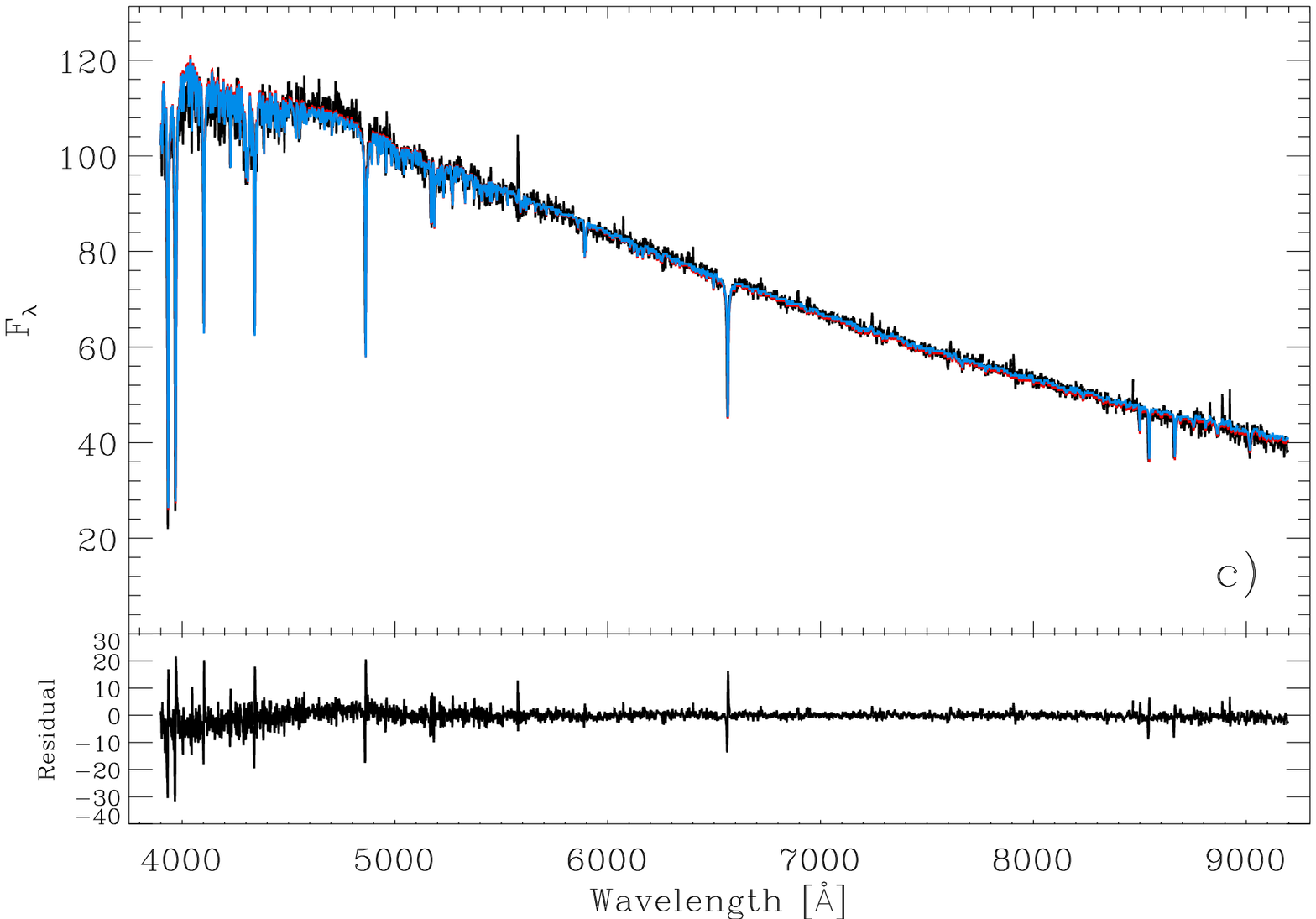}
\includegraphics[width=0.49\columnwidth]{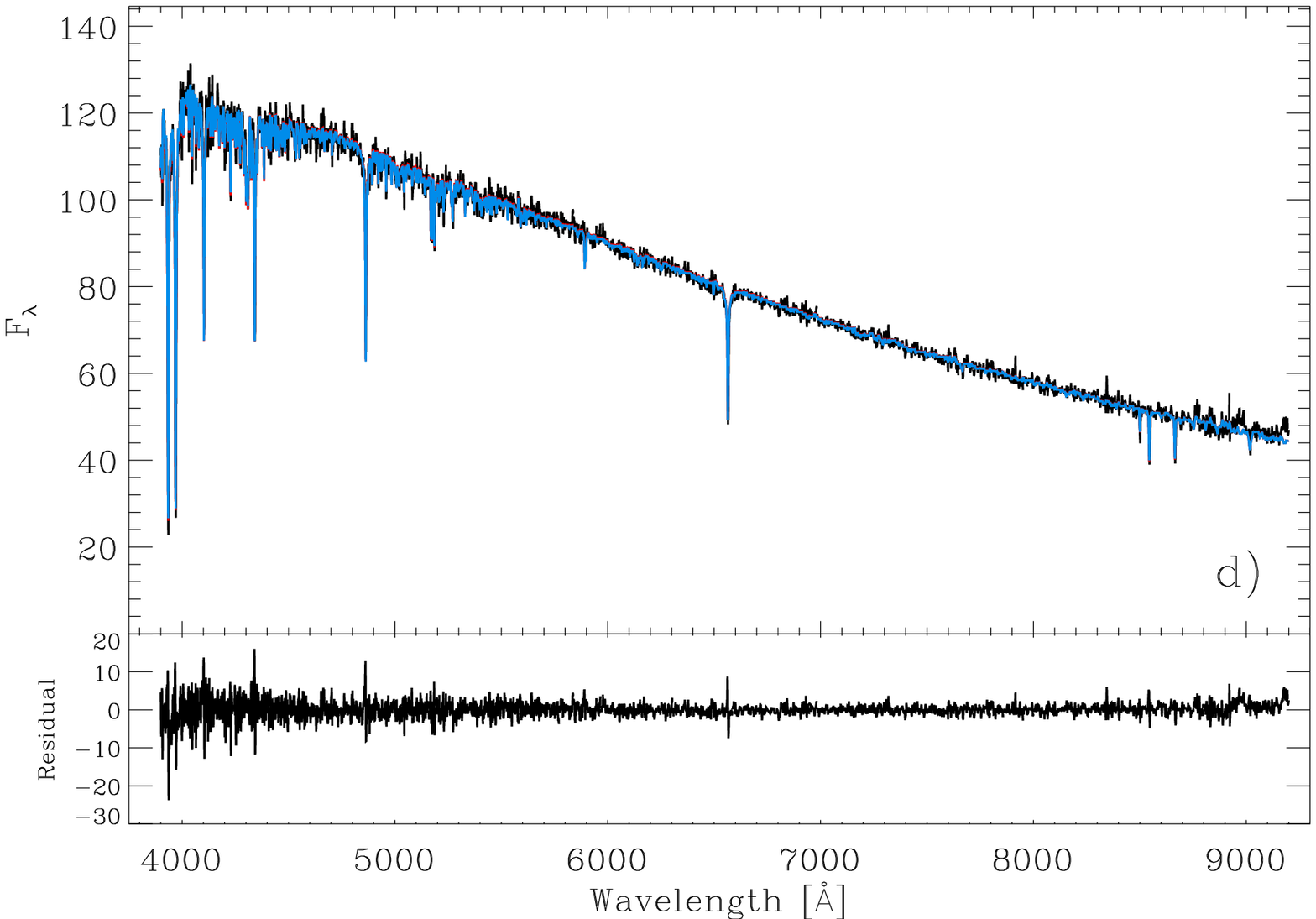}
\caption{Four examples of the spectrum reconstruction using simulated magnitudes in
filters $g$, $r$ and $i$. The original spectrum is shown in black. The spectrum
reconstructed from the exact projections along the first 3 principal components of \cite{mcgurk10}
is shown in red. This constitutes the best reconstruction of the spectrum we can
achieve with our method. The spectrum reconstructed using our scheme is shown in blue, with the
residual indicated in the lower subpanel of each panel. Note the
similarity between the red and blue curves.}
\label{fig:reconstruction_example}
\end{figure*}

We carry out reconstructions using magnitudes synthesized from observed spectra included in SDSS Data Release 7. 
The synthetic magnitudes are obtained following Eq. (\ref{eq:magnitudes}) for 
filters $g$, $r$ and $i$. These could be taken directly from the photometric observations instead.
The sensing matrix is built using Eq. (\ref{eq:sensing_matrix_final}).  
The spectrum of four representative stars from the sample are reconstructed solving the linear
system of Eq. (\ref{eq:linear_system_reconstruction}). The success of the technique
is shown in Fig. \ref{fig:reconstruction_example}. As stated before, to this aim we make use 
only of the mean spectrum and two principal components. The original noisy spectrum is shown in black 
color. The projection of the observed spectrum on the
space spanned by the first three principal components is shown in red color. This constitutes
the best possible reconstruction that we could achieve with the presented method.
The spectrum reconstructed with our technique
using the magnitude in the filters $g$, $r$, and $i$ is shown in blue. Note that the reconstruction
closely follows the red curve, indicating that a good reconstruction is possible and that the
projection along the first three principal components can be obtained reliably
from the linear measurements made with the SDSS filters. 

The fundamental characteristics of the spectra
are reproduced with precision, making it possible to empirically infer spectroscopic 
quantities using photometric measurements. At the same time, thanks to the projection
along the principal components, the spectrum reconstructed from $gri$ magnitudes is
automatically denoised \citep[see, e.g.,][for the denoising capabilities of PCA]{marian_pcafilter08}.
Particularly large residuals are visible at specific wavelengths in
Fig. \ref{fig:reconstruction_example}. In panel a) we can spot some issues with sky
removal at the green [OI] line (5577 \AA) and
in the IR end. In panels b), c) and d) there are strong
residuals around the transitions of the Balmer series and
other strong features. These residuals change sign
very quickly around the central wavelength, signaling
a horizontal offset between the original and the
reconstructed spectra, most likely related to the
Dopper velocity shifts, which have not been corrected but
happen to be small enough for the star depicted in panel a).

\begin{figure*}[!t]
\centering
\includegraphics[width=0.49\columnwidth]{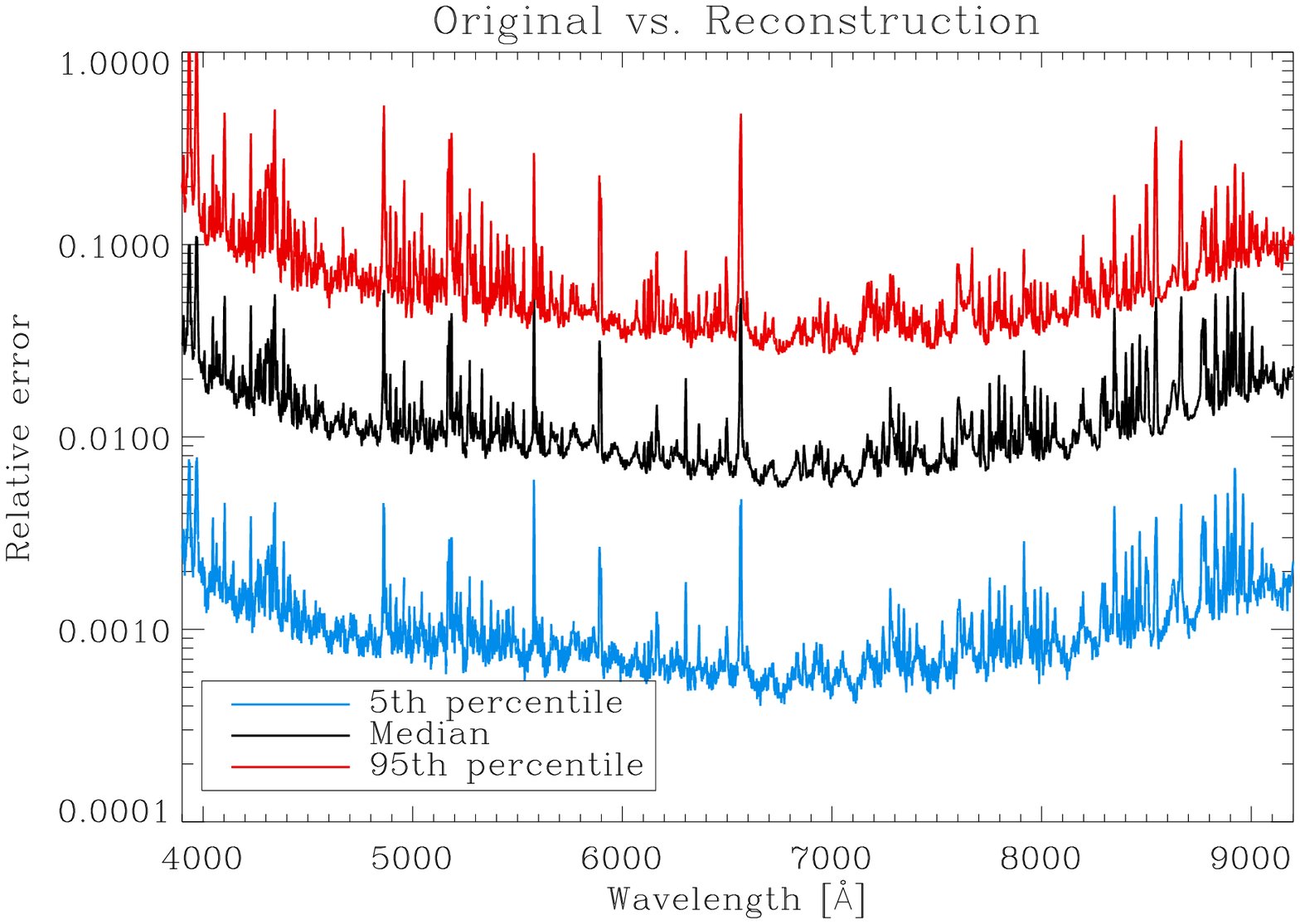}
\includegraphics[width=0.49\columnwidth]{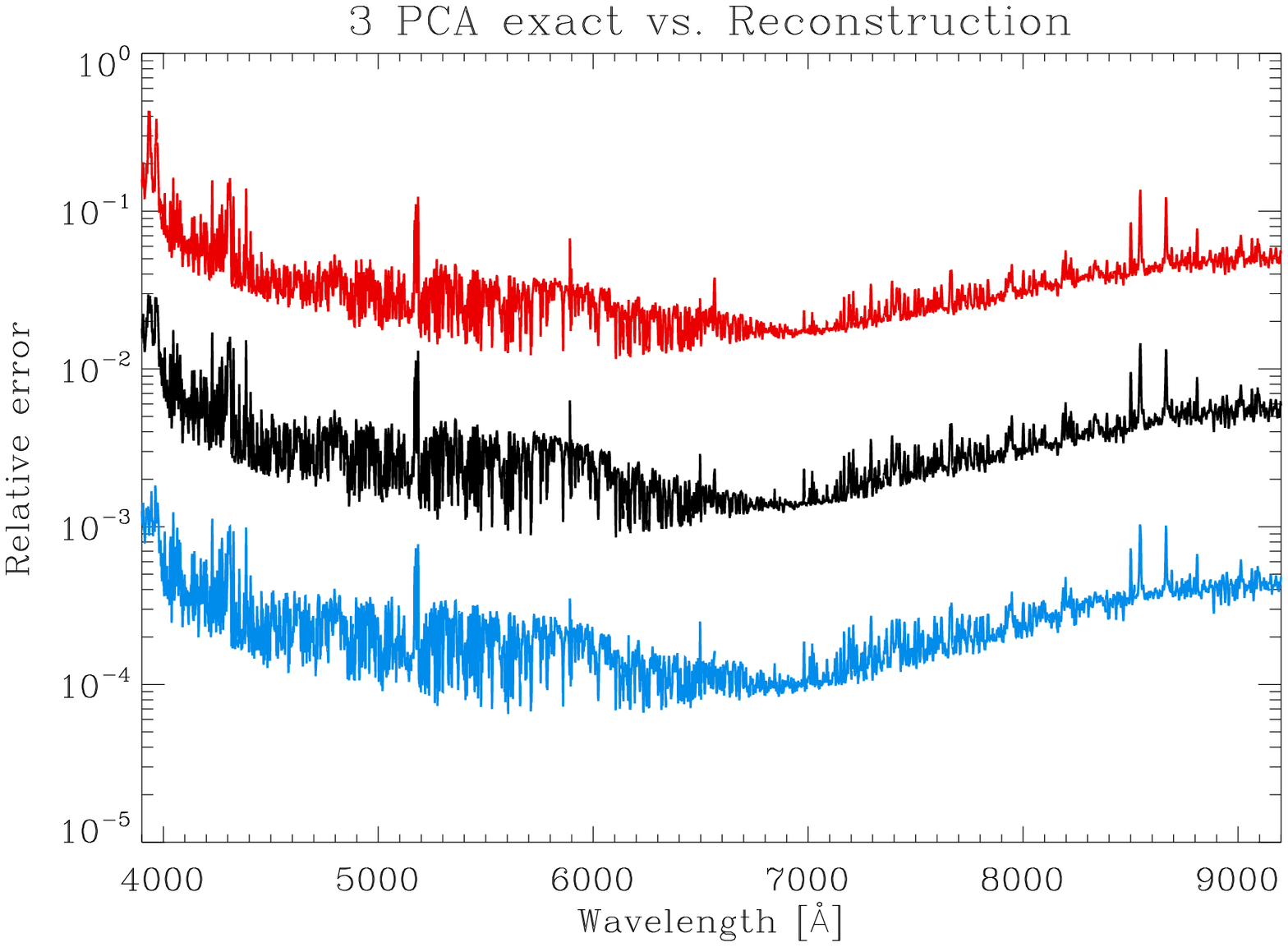}
\includegraphics[width=0.49\columnwidth]{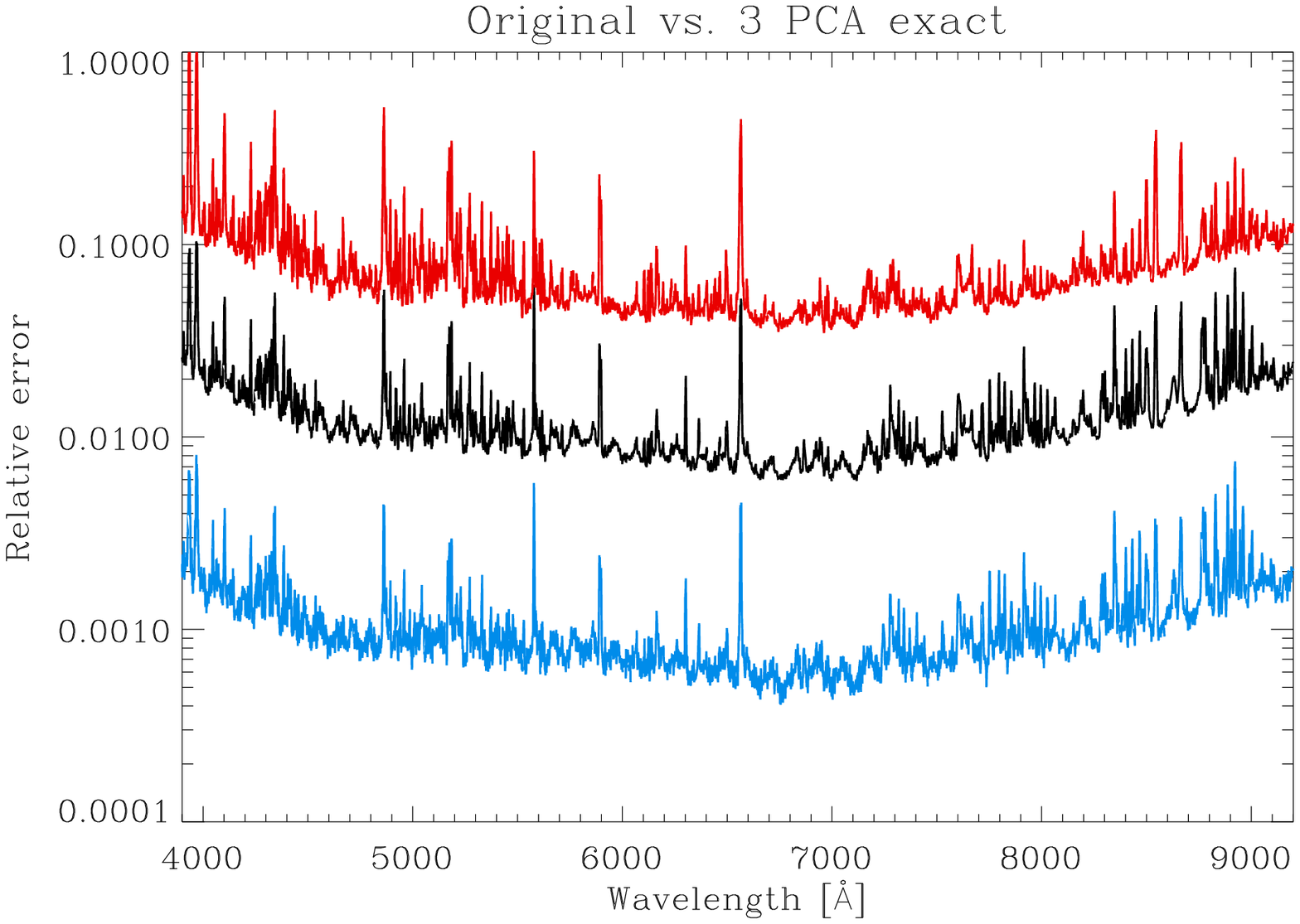}
\includegraphics[width=0.49\columnwidth]{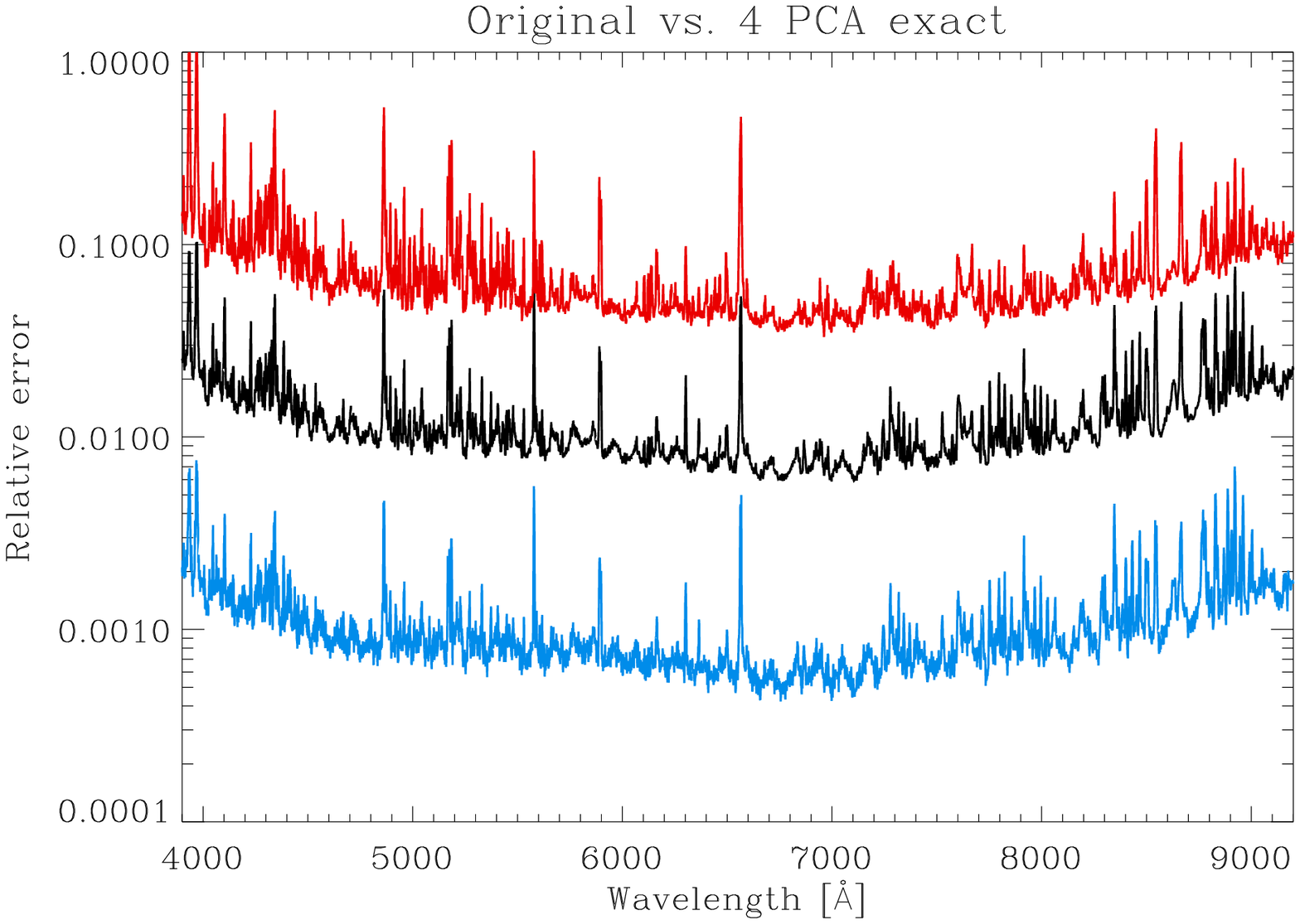}
\caption{Statistical comparison showing the viability of the reconstruction scheme for
5000 stars chosen at random from the database of \cite{mcgurk10}. We show the 5th (blue curve), 
50th (black curve) and 95th (red curve) percentiles of the relative error distribution for each 
wavelength. The upper left panel presents the relative error obtained between the reconstructed
spectrum using our method and the original noisy spectrum. It is clear that 50\% of the stars have relative
errors below 1-2\% while relative errors are smaller than 10\% with 95\% probability. The 
upper right panel shows the relative error between the reconstruction using photometric
data and the denoised spectrum computed with the exact projections along the first three principal components. For
reference, the lower panels show the comparison between the original spectra and the
ones reconstructed using the exact projections along the first three (lower left panel) and
the first four (lower right panel) principal components.}
\label{fig:statistical_difference}
\end{figure*}

As four examples are not statistically relevant, we have carried out an analysis of the
differences between the reconstructed and original spectra for 5000 stars chosen at random
from the seventh data release of the SDSS \citep{abazajian09}.
We have verified that a sample of this size chosen at random covers all $g-r$ bins and provides
reliable statistics. The quality 
is characterized, at each wavelength, by the 5th, 50th and 95th percentile
of the distribution of relative errors from the reconstructed spectrum and the original one.
The results are shown in the upper-left panel of Fig. \ref{fig:statistical_difference}. As stated, the
50th percentile (the median) is represented as a black curve and indicates that half of the stars 
can be reconstructed with relative errors smaller than 2\% (from $\sim$4200 to 9000 \AA). 
The 5th and 95th
percentiles (blue and red curves) are also indicated in the upper left panel
of Fig. \ref{fig:statistical_difference}.
It is demonstrated that the reconstructions can be done with relative errors well below 10\% for
95\% of the stars. Of course, this does not rule out the presence of 5\% of the stars
with relative reconstruction errors potentially larger than 10\%.

For reference, in Fig. \ref{fig:statistical_difference} reconstructions with the 
first three (lower left panel) and four (lower right panel) principal components
as obtained from \cite{mcgurk10} are compared with the original spectra. These plots
summarize the quality of the PCA recreations. Although PCA reconstructions with 
relative errors below or of the order of 1\% are possible for 50\% of the stars, a fraction of 
stellar spectra will incur (even knowing exactly the projections along the principal
components) relative errors larger than 10\%. Note also that the improvement
on the reconstruction is marginal when using four instead of three principal components.
An indication of the quality of the PCA reconstruction is that the
difference between the initial magnitudes and the reconstructed magnitudes has 
a standard deviation of 0.008 for $g$, $r$ and $i$ when using 4 principal components.
When using only one principal component, this number increases up to 0.03 for $g$ and $r$
and to 0.05 for $i$.

It is important to realize that the residuals shown
in Fig. \ref{fig:statistical_difference} are significantly higher for wavelengths
with lines than in continuum regions. This suggests
that the PCA reconstruction is reproducing well
the continuum shape, but not so the lines' strength.
However, lines are crucial for recovering
information on surface gravity and chemical composition.
A possibility  for improvement is therefore to
perform PCA on continuum-corrected spectra.

If instead of using the original spectra as reference, reconstructions are 
compared with the spectra projected onto the space spanned by the
first three principal components, the results are those shown in the upper right panel of
Fig. \ref{fig:statistical_difference}. It is clear from this plot that our method is
able to reliably extract the projection along the first three principal components from 
photometric information. 

\begin{figure*}[!t]
\centering
\includegraphics[width=0.49\columnwidth]{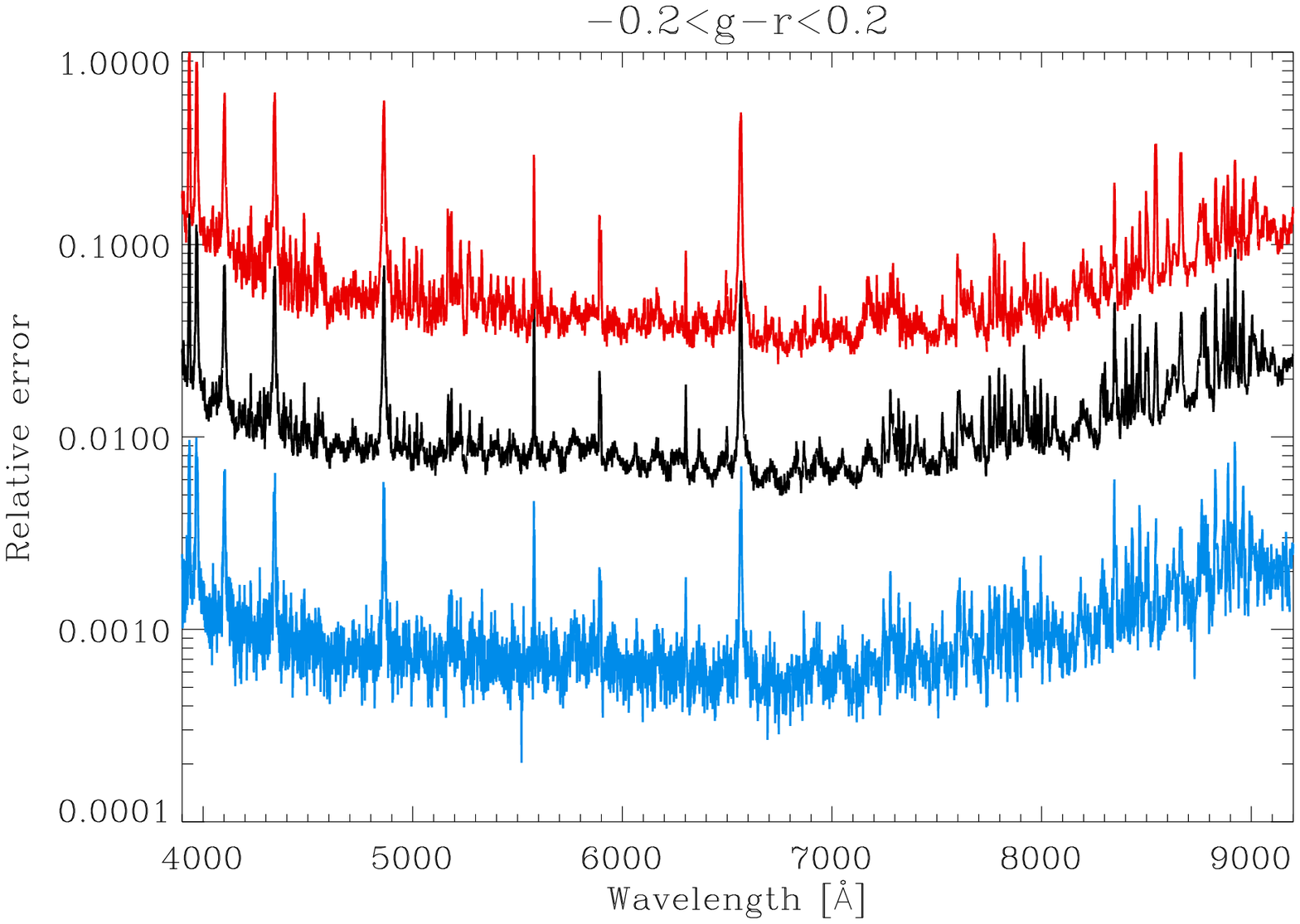}
\includegraphics[width=0.49\columnwidth]{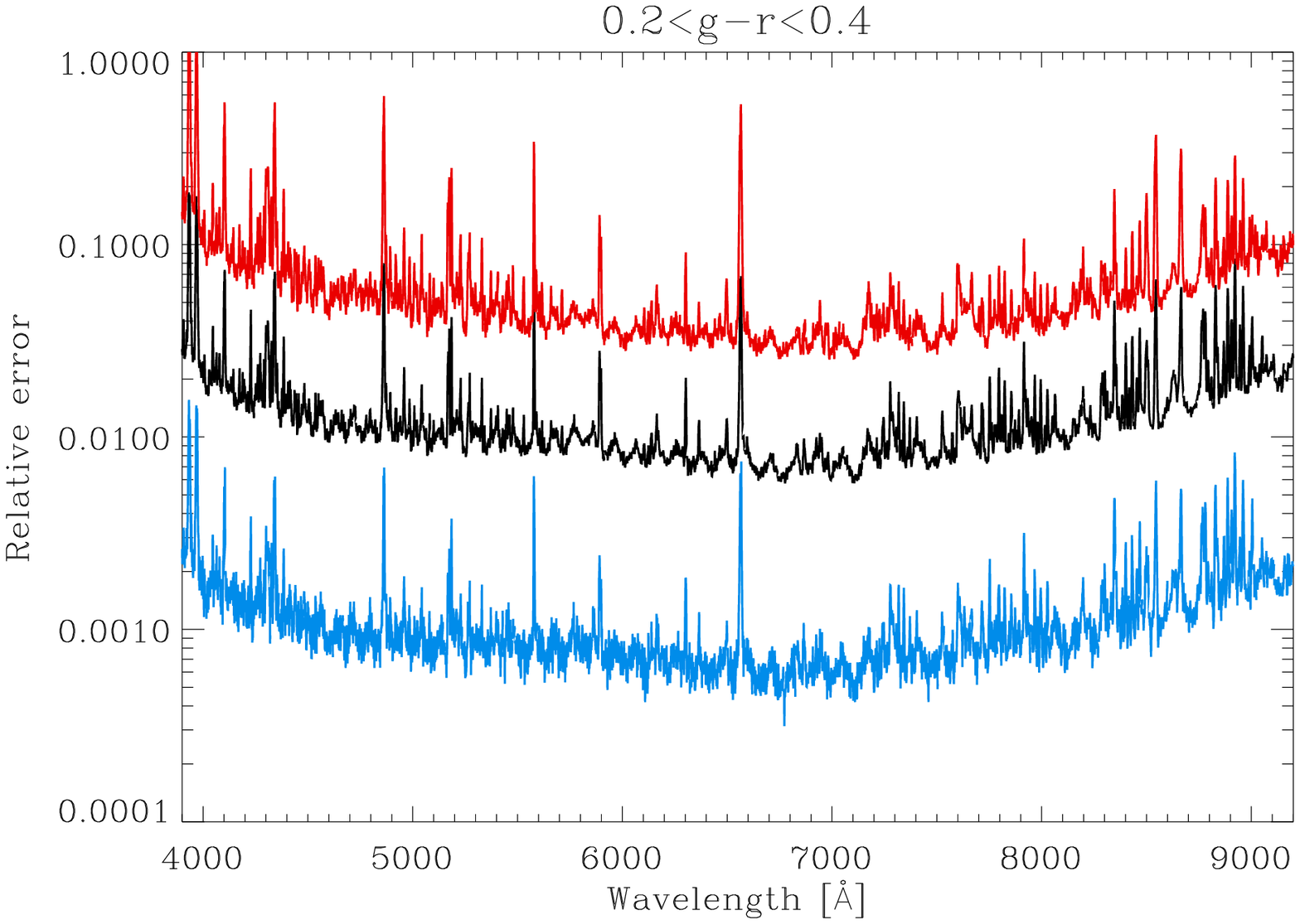}
\includegraphics[width=0.49\columnwidth]{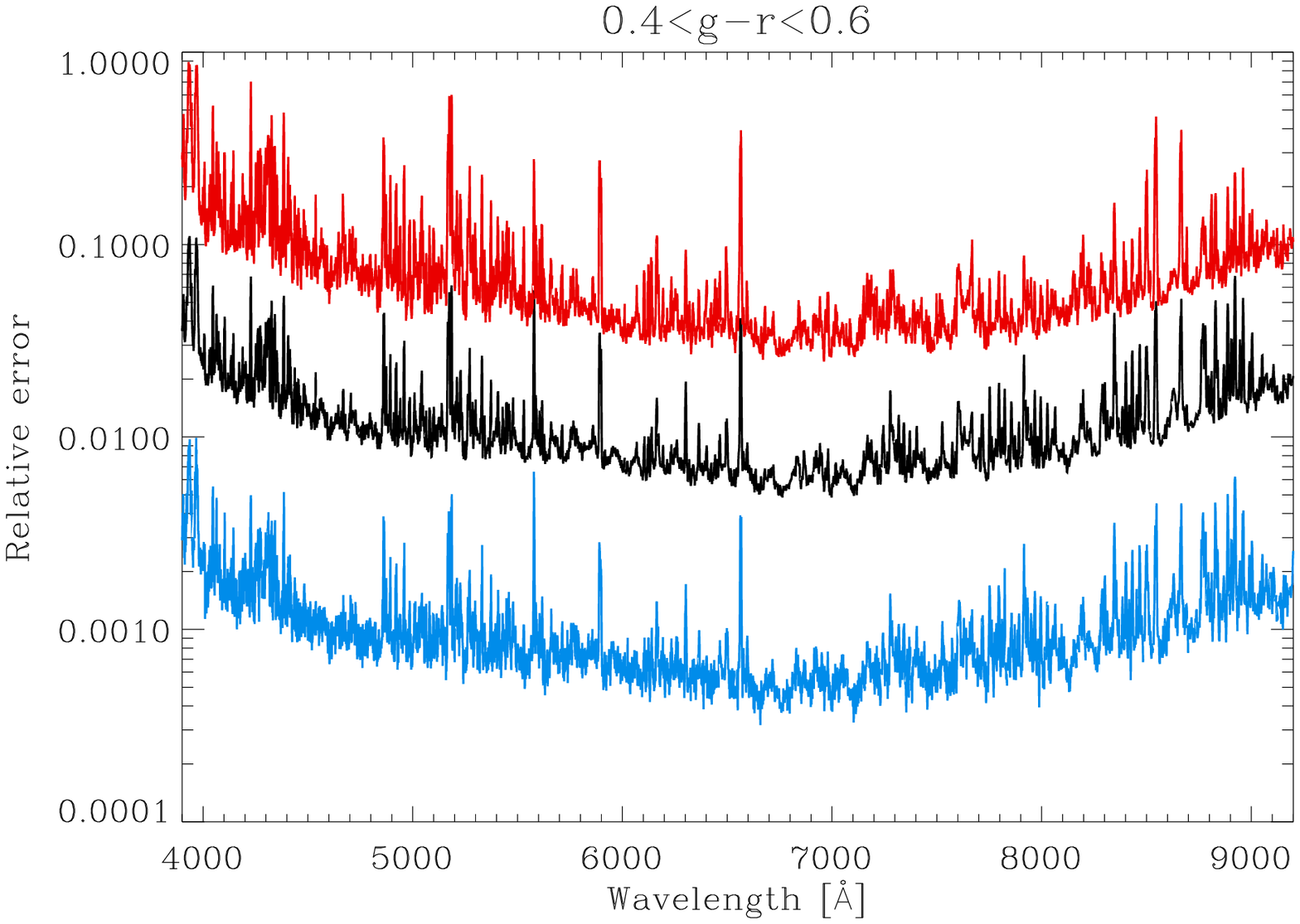}
\includegraphics[width=0.49\columnwidth]{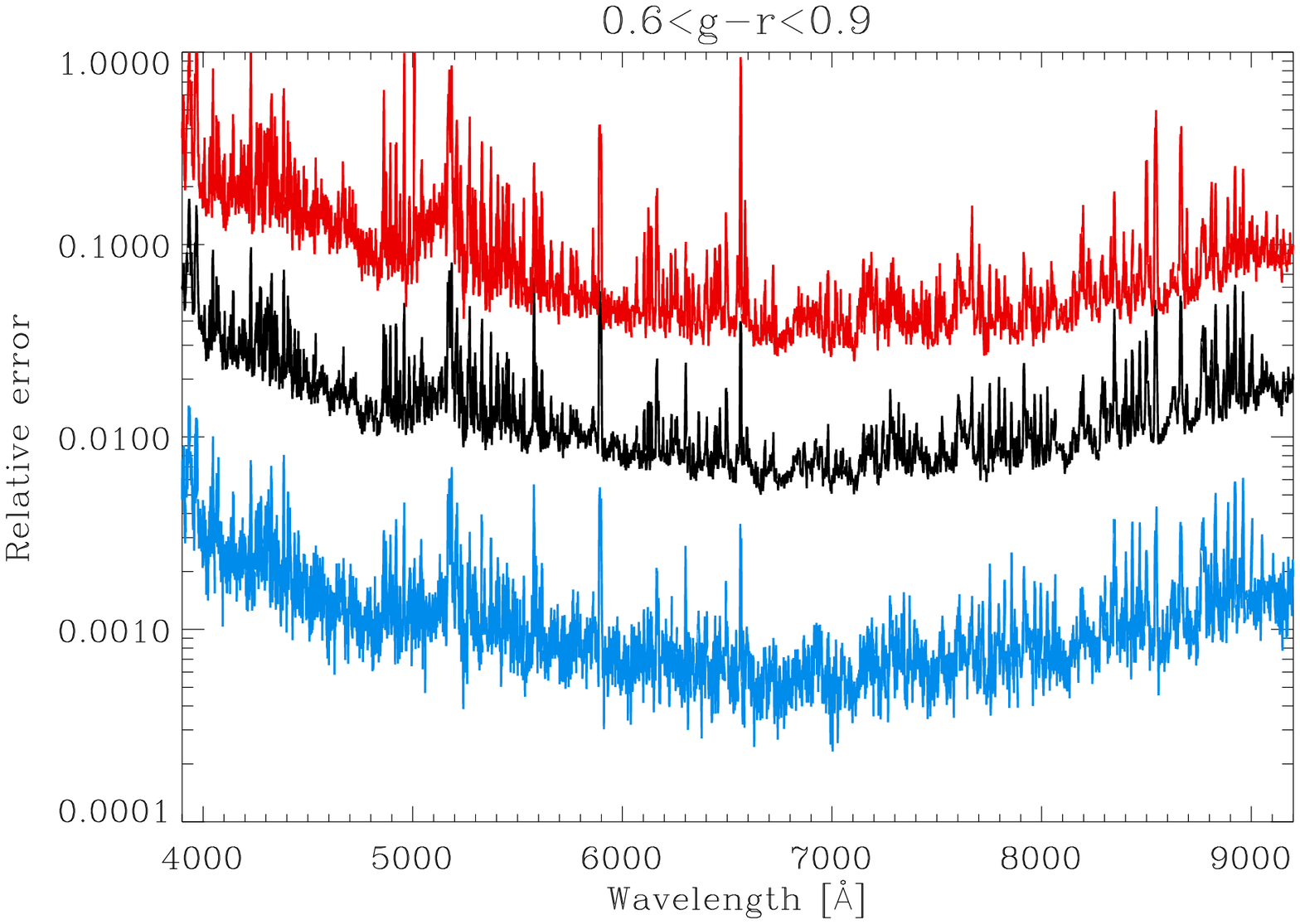}
\caption{Relative error between the reconstructed spectrum using our method and the 
original noisy spectrum for the sample separated in color bins. Note that reconstructions of 
stars with higher effective
temperatures (smaller $g-r$) are of better quality due to the small number
of spectral features. On the contrary, cooler stars (larger $g-r$) tend to have
molecular bands that difficult the reconstruction. In any case, even in the
least favourable case, 50\% of the stars have relative
errors below 4-5\% while relative errors are smaller than 20\% with 95\% probability (with a
large wavelength region with errors below 10\% for 95\% of the stars).
The colors have the same meaning as those of Fig. \ref{fig:statistical_difference}.}
\label{fig:statistical_difference_bins}
\end{figure*}

In order to analyze the quality of reconstructions as a function of the spectral
type, we show in Fig. \ref{fig:statistical_difference_bins} the relative
error between the reconstruction and the original noisy spectrum for stars
in different bins of $g-r$. As expected, we note that our reconstruction
leads to slightly worse results in cooler stars. This is a consequence of
the fact that the spectrum of cool stars is relatively more complex and the photometry
is not able to capture their full variation. In any case, even in the
less favorable case for stars with $g-r > 0.6$ ($T_\mathrm{eff} \lesssim 5000$ K),
reconstructions are below 10\% for 95\% of the stars in a large wavelength
range.


\section{Influence of errors}
\label{sec:influence_errors}
Observed magnitudes are always inherently accompanied by an error bar. It is important to
quantify the effect of this error on the reconstruction of the spectrum. Assuming Gaussian
errors, an error bar of standard deviation $\sigma_{m_j}$ in magnitudes at filter $j$ translates
into an error bar in the flux at the same filter of:
\begin{equation}
\label{eq:error_o}
\sigma_{o_j} = 0.4 \ln 10 o_j \sigma_{m_j} \approx 0.92 o_j \sigma_{m_j}.
\end{equation}
Even if we assume that the error bars of the observations are not correlated, the
resulting error bars for the projections along the mean spectrum and the principal
components are correlated. Assuming that the matrix $\mathbf{G}$ is noise-free\footnote{This
implies assuming that the filter and atmospheric transmissions are known
with absolute certainty. Obviously, this can be relaxed without too much effort, although
the final expression for the covariance matrix of the projection along the principal
components contains another contribution due to the uncertainties in the $\mathbf{G}$ matrix 
\citep[see][for an example in another field]{asensio_appopt08}.}, error 
propagation in the solution of the linear system of Eq. (\ref{eq:linear_system_reconstruction}) leads 
to the following formula for the covariance matrix of the projection along
the principal components:
\begin{equation}
\mathbf{C}_a = \mathbf{G}^{-1} \mathbf{C}_b (\mathbf{G}^{-1})^\dag,
\end{equation}
where
\begin{equation}
\mathbf{C}_b = \mathbf{T} \mathbf{C}_o \mathbf{T}^\dag,
\end{equation}
where $T_{ij}=t_i^j / \sigma^2_{o_j}$.
For simplicity, we assume that the correlation matrix of the observed fluxes is diagonal
and the diagonal elements are computed from Eq. (\ref{eq:error_o}).
Finally, the covariance matrix for the reconstructed spectrum is given by:
\begin{equation}
\mathbf{C}_F = \mathbf{W}^\dag \mathbf{C}_a \mathbf{W}.
\end{equation}

\begin{figure}[!t]
\centering
\plotone{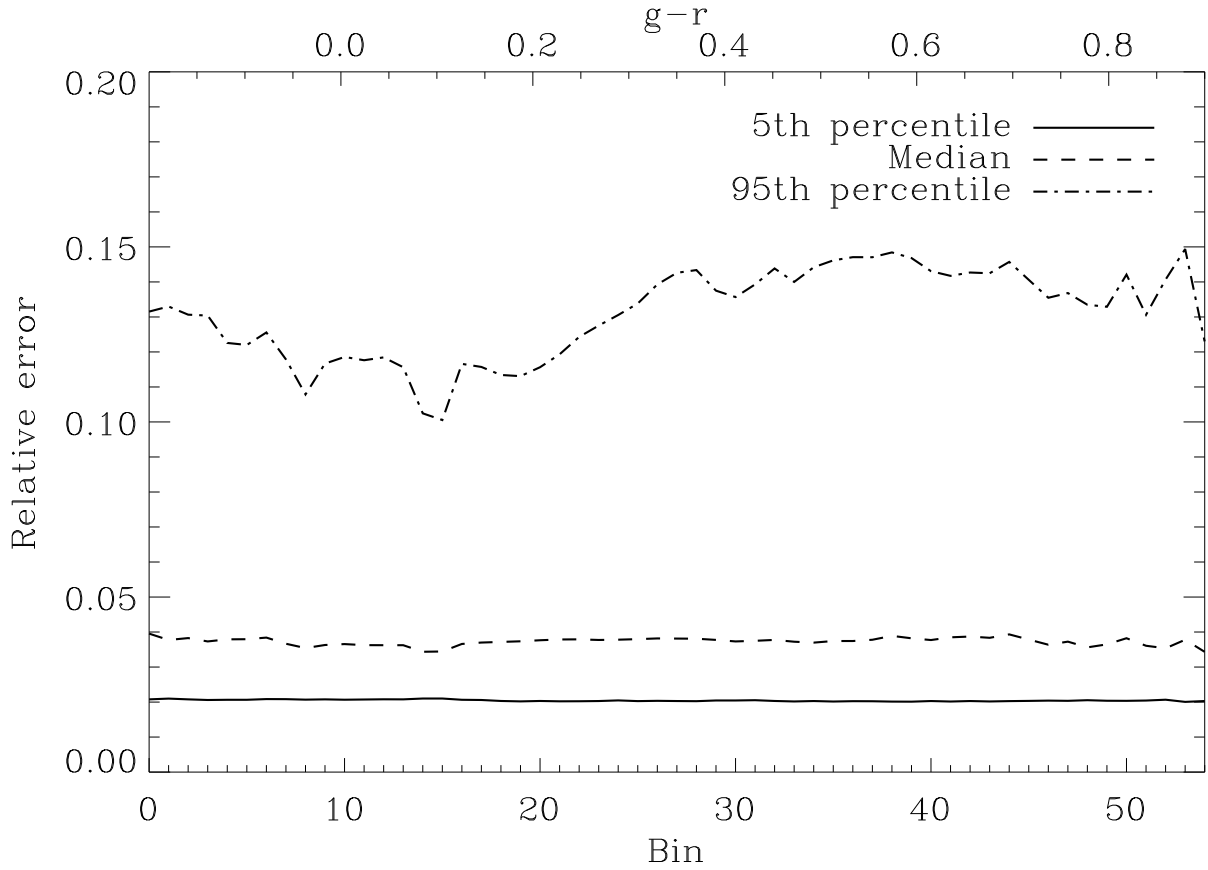}
\caption{Standard deviation of the error in the reconstruction of the projection of the
spectrum along the mean spectrum and the first two principal components for each
bin. The color $g-r$ for each bin is also indicated in the upper axis.
The error is normalized to the quantity $\sigma_m$ and estimated using the
average flux in each filter for each bin.}
\label{fig:error_magnitude}
\end{figure}

It is difficult to characterize the sensitivity to errors in the observed
magnitudes because of the large variability in the stellar fluxes.
For presentation purposes and to give a rough estimation, let us assume that 
all Sloan magnitudes have $\sigma_m=0.03$ mag, which
is representative of more than 95\% of the observed stars. Likewise, let us pick a 
representative value for the flux at each filter $o_j$ as the average in each bin.
Following the previous expressions, we show in Figure \ref{fig:error_magnitude} the 
standard deviation of the error in the
projection along the principal components (equivalent to the diagonal elements 
$\mathrm{cov}(a_i,a_i)$) normalized to the product of the observational error in the magnitude and 
the mean projection along the mean spectrum. It has been estimated for the average flux in each bin. The 
results indicate that the relative error in the projection along the principal components is
roughly similar to $\sigma_m$. In the SDSS database, typical errors range
from 0.01 to 0.05 mag, with more than 95\% of the stars with errors less than
0.03 mag. Therefore, relative errors of 1-3\% are induced in the reconstruction due to
the presence of uncertainties in the observed magnitudes.


The principal components of \cite{mcgurk10} have been computed by shifting all spectra to
a zero radial velocity common wavelength axis. Therefore, the reconstructions we carry
out give as an output the spectrum at zero radial velocity and contain no information
on radial velocities. However, the fluxes measured
photometrically with the $gri$ filters contain the influence of the radial velocity. This effect
cannot be compensated for and it is not clear whether this might have an influence
on the final reconstruction. From SDSS data, the distribution of radial velocities 
induce Doppler shifts that are, with $\sim$95\% probability, smaller than 6 pixels ($\sim$414 km/s). According to
the reconstructions shown in Figs. \ref{fig:reconstruction_example} and 
\ref{fig:statistical_difference}, that
were performed with the original spectrum, Doppler shifts increase
errors in the lines, as suggested by the antisymmetric residuals
in the strongest lines of panels b) c) and d) in Fig. \ref{fig:reconstruction_example}
Such errors are masked in Fig. \ref{fig:statistical_difference} by the symmetrization induced by
averaging out over many stars. Therefore, although radial velocities cannot be estimated with
our method, its effect on the quality of the reconstruction
is not very important, except for spectral lines.

\section{Conclusions}
We have presented a method to reconstruct stellar spectra from SDSS photometry that is
fast, efficient and reliable. Although it might sound magical, this reconstruction is 
made possible thanks to the sparsity of stellar spectra, which can be expressed as
a linear combination of a few eigenspectra obtained from a principal component
decomposition. Our method returns the projection of the observed spectrum 
onto the space spanned by the first three principal components just from the photometric
$g$, $r$ and $i$ magnitudes. As a consequence, the resulting spectrum is
simultaneously denoised and reconstructed. We have analyzed the statistical properties of 
the regenerated spectra and verified that the residuals are roughly compatible with the noise
present in the observations, albeit they are not random. We have also analyzed the influence of observational errors in
the magnitudes and the presence of non-zero radial velocities on the reconstruction. Both
of them produce very small effects on the performance of our algorithm.

Recently, \cite{mcgurk10} has investigated the possible correlation between 
stellar parameters and the projections of the spectrum along the principal
components. Since PCA is a linear technique and the stellar parameters are
typically nonlinear combinations of parts of the observed spectrum, a strong
correlation is not to be expected, as shown by \cite{mcgurk10}. In our case though,
we are able to reconstruct the full spectrum from photometric observables. This
opens up the possibility of applying standard techniques for inferring stellar
parameters to the reconstructed spectrum.

\begin{acknowledgements}
Financial support by the Spanish Ministry of Science and Innovation through project AYA2007-63881 is gratefully acknowledged.

Funding for the SDSS and SDSS-II has been provided by the Alfred P. Sloan Foundation, the 
Participating Institutions, the National Science Foundation, the U.S. Department of Energy, 
the National Aeronautics and Space Administration, the Japanese Monbukagakusho, the Max 
Planck Society, and the Higher Education Funding Council for England. The SDSS Web Site 
is http://www.sdss.org/.

The SDSS is managed by the Astrophysical Research Consortium for the Participating 
Institutions. The Participating Institutions are the American Museum of Natural History, 
Astrophysical Institute Potsdam, University of Basel, University of Cambridge, Case Western 
Reserve University, University of Chicago, Drexel University, Fermilab, the Institute for 
Advanced Study, the Japan Participation Group, Johns Hopkins University, the Joint Institute 
for Nuclear Astrophysics, the Kavli Institute for Particle Astrophysics and Cosmology, the 
Korean Scientist Group, the Chinese Academy of Sciences (LAMOST), Los Alamos National 
Laboratory, the Max-Planck-Institute for Astronomy (MPIA), the Max-Planck-Institute for 
Astrophysics (MPA), New Mexico State University, Ohio State University, University of 
Pittsburgh, University of Portsmouth, Princeton University, the United States Naval 
Observatory, and the University of Washington.
\end{acknowledgements}


\clearpage

\end{document}